\documentclass[journal=nanolett ,manuscript=article]{achemso}

\usepackage[version=3]{mhchem}
\usepackage{natbib}

\setkeys{acs}{keywords = true}
\def\acs@type@default{letter}
\def\acs@type@list{letter}
\SectionNumbersOff

\author{Sergey A. Nikolaev}
\email{nikolaev.s.es@osaka-u.ac.jp}
\affiliation{The University of Osaka, Toyonaka, Osaka 560-8531, Japan}

\author{Mairbek Chshiev}
\affiliation{Univ. Grenoble Alpes, CEA, CNRS, Spintec, 38000, Grenoble, France}
\alsoaffiliation{Institut Universitaire de France, 75231 Paris, France}

\author{Fatima Ibrahim}
\affiliation{Univ. Grenoble Alpes, CEA, CNRS, Spintec, 38000, Grenoble, France}

\author{Sachin Krishnia}
\affiliation{Laboratoire Albert Fert, CNRS, Thales, Universit\'e Paris-Saclay, 91767 Palaiseau, France}

\author{Nicolas Sebe}
\affiliation{Laboratoire Albert Fert, CNRS, Thales, Universit\'e Paris-Saclay, 91767 Palaiseau, France}

\author{Jean-Marie George}
\affiliation{Laboratoire Albert Fert, CNRS, Thales, Universit\'e Paris-Saclay, 91767 Palaiseau, France}

\author{Vincent Cros}
\affiliation{Laboratoire Albert Fert, CNRS, Thales, Universit\'e Paris-Saclay, 91767 Palaiseau, France}

\author{Henri Jaffr\`es}
\affiliation{Laboratoire Albert Fert, CNRS, Thales, Universit\'e Paris-Saclay, 91767 Palaiseau, France}

\author{Albert Fert}
\affiliation{Laboratoire Albert Fert, CNRS, Thales, Universit\'e Paris-Saclay, 91767 Palaiseau, France}

\title{Large Chiral Orbital Texture and Orbital Edelstein Effect in Co/Al Heterostructure}

\begin{document}

\begin{abstract}
Recent experiments by S. Krishnia et al., Nano Lett. 23, 6785 (2023) reported an unprecedentedly large enhancement of torques upon inserting thin Al layer in Co/Pt heterostructure that suggested the presence of a Rashba-like interaction at the metallic Co/Al interface. Based on first-principles calculations, we reveal the emergence of a large helical orbital texture in reciprocal space at the interfacial Co layer, whose origin is attributed to the orbital Rashba effect due to the formation of the surface states at the Co/Al interface and where spin-orbit coupling is found to produce smaller contributions with a higher-order winding of the orbital momentum. Our results unveil that the orbital texture gives rise to a non-equilibrium orbital accumulation producing large current-induced torques, thus providing an essential theoretical background for the experimental data and advancing the use of orbital transport phenomena in all-metallic magnetic systems with light elements.  
\end{abstract}

\par Search for efficient ways to control the magnetization in magnetic materials has been one of the central activities in the field of spintronics, which exploits the intrinsic spin of an electron with a focus on providing novel functionalities in electronic devices. In this regard, spin-orbit coupling (SOC) has brought a progressive venue to achieving high-efficiency electrical control of magnetization via spin-orbit torques (SOT),\cite{RevModPhys.91.035004} which historically relied on two cornerstone mechanisms. On the one hand, the spin Hall effect in nonmagnetic materials with strong SOC, such as Pt, allows generating spin currents in response to the flowing charge current.\cite{1971ZhPmR..13..657D,PhysRevLett.83.1834,PhysRevLett.94.047204,RevModPhys.87.1213} On the other hand, spatial inversion symmetry breaking at the interface can give rise to the spin Rashba-Edelstein effect when a non-equilibrium spin accumulation $\delta\boldsymbol{S}$ is produced by an electric field as a result of the interplay of low dimensionality and SOC.\cite{Bychkov1984Jan,EDELSTEIN1990233} Both the spin currents, when injected into adjacent magnetic layers, and the local spin accumulation induced at the interface of a magnetic heterostructure can exert a SOT on the local magnetic moments $\boldsymbol{M}$, which is conventionally characterized in terms of the field-like $\boldsymbol{\tau}_{\mathrm{FL}}\sim\boldsymbol{m}\times\delta\boldsymbol{S}$ and damping-like $\boldsymbol{\tau}_{\mathrm{DL}}\sim\boldsymbol{m}\times(\boldsymbol{m}\times\delta\boldsymbol{S})$ components with $\boldsymbol{m}=\boldsymbol{M}/|\boldsymbol{M}|$.\cite{Garello2013Aug} The spin Hall effect is considered the primary source of $\boldsymbol{\tau}_{\mathrm{DL}}$ in metallic systems,\cite{Liu2012May,Fan2014Jan} while $\boldsymbol{\tau}_{\mathrm{FL}}$ is predominantly associated with the Rashba-Edelstein effect at the ferromagnet/oxide interfaces.\cite{MihaiMiron2010Mar} In general, both effects act in concert producing SOTs, whose competition governs magnetization dynamics and upon which reversible magnetization switching\cite{MihaiMiron2010Mar,Miron2011Aug,PhysRevLett.106.036601,PhysRevLett.109.096602} and domain wall motion\cite{Miron2011Jun,Thiaville2012Dec,Ryu2013Jul,PhysRevB.87.020402} can be achieved in a very efficient way. 
 
\par Over the recent years, there has been a lof of evidence that the spin Hall and Rashba-Edelstein effects may not be the only principal ways to generate SOTs. It was shown theoretically that, in addition to the spin accumulation, a flowing charge current can induce a non-equilibrium orbital accumulation, and, similar to the spin Hall effect, one can realize electrical generation of the transverse orbital currents, or the orbital Hall effect, which is in many cases predicted to exceed its spin counterpart.\cite{PhysRevB.77.165117,PhysRevLett.100.096601} In the same vein, the breaking of spatial inversion symmetry at the interface can result in non-trivial orbital textures in the momentum space regardless of the presence of SOC and the so-called orbital Rashba effect\cite{PhysRevLett.107.156803,PhysRevB.85.195401,PhysRevB.85.195402,PhysRevB.87.041301,Go2017Apr,PhysRevB.103.L121113} which can produce a substantial orbital accumulation under applied electric fields due to the orbital Edelstein effect.\cite{Yoda2015Jul,Yoda2018Feb} Importantly, several studies established that SOTs can be partly mediated by orbital angular momentum and that the injection of orbital currents into a ferromagnet can excite magnetization dynamics.~\cite{PhysRevResearch.2.013177,PhysRevResearch.2.033401} Nowadays, orbitronics is an emerging field of research that aims at exploiting the transport of orbital angular momentum and explores possibilities for generating and manipulating the orbital currents, either alongside or independently of the spin degrees of freedom, which spurred an ongoing experimental endeavour on direct observation of the orbital transport phenomena.\cite{Lee2021Nov,PhysRevLett.128.067201,Hayashi2023Feb,Choi2023Jul} In the context of magnetic torques, recent experimental studies revealed large enhancement of SOTs in thulium iron garnet TmIG/Pt/CuO$_{x}$\cite{PhysRevLett.125.177201} and CoFe/Cu/Al$_{2}$O$_{3}$\cite{PhysRevB.103.L020407} heterostructures which were proposed to be of the orbital origin. 

\par Our recent experiments demonstrated strong enhancement of the current-induced torques in ultrathin Co ferromagnet by insertion of light metal elements.\cite{acs.nanolett.2c05091} In particular, an unprecedentedly large ratio of the field-like and damping-like torques has been observed upon adding thin Al layer to Co/Pt heterostructure. With 3 nm of Al, $\boldsymbol{\tau}_{\mathrm{FL}}$ and $\boldsymbol{\tau}_{\mathrm{DL}}$ are increased up to a factor of 11 and 5, respectively, as compared to the same stacking without Al, while similar stackings with Cu inserted instead of Al are found to have negligible effects on the torques. Such giant enhancement of the current-induced torques by introduction of Al layer adjacent to Co and the predominance of the FL torque over the DL torque support the interfacial origin of SOTs and the existence of a large Rashba-type interaction, which is, in turn, not expected for a light-metal interface. Yet, since SOC is small in light elements such as Al, the induced SOTs cannot be directly ascribed to conventional spin Hall and Rashba-Edelstein effects. Despite a number of pioneering studies uncovering the theory of orbital torques, the microscopic origin of enhanced current-induced torques in these systems remains to be understood.

\par In the present study, we provide a detailed theoretical evidence on the existence of a large chiral orbital texture at the Co/Al interface and the orbital origin of SOTs in Co/Al heterostructure. We carry out first-principles calculations for Co(0001)/Cu(111) and Co(0001)/Al(111) heterostructures with a single interface and vacuum region by varying the number of constituent layers and considering several possible stackings at the interface. Electronic structure calculations are performed within density functional theory and generalized gradient approximation\cite{PhysRevLett.77.3865} (GGA), as implemented in the Vienna Ab-initio Simulation Package.\cite{PhysRevB.54.11169,PhysRevB.59.1758,PhysRevB.13.5188} All calculations consider the ferromagnetic state, with an easy axis perpedicular to the interface in the presence of SOC, and structural optimization is carried out for each considered heterostructure. Further details are given in Section~I of Supplementary Material. 

\par
\begin{figure}
\includegraphics[width=1.0\textwidth]{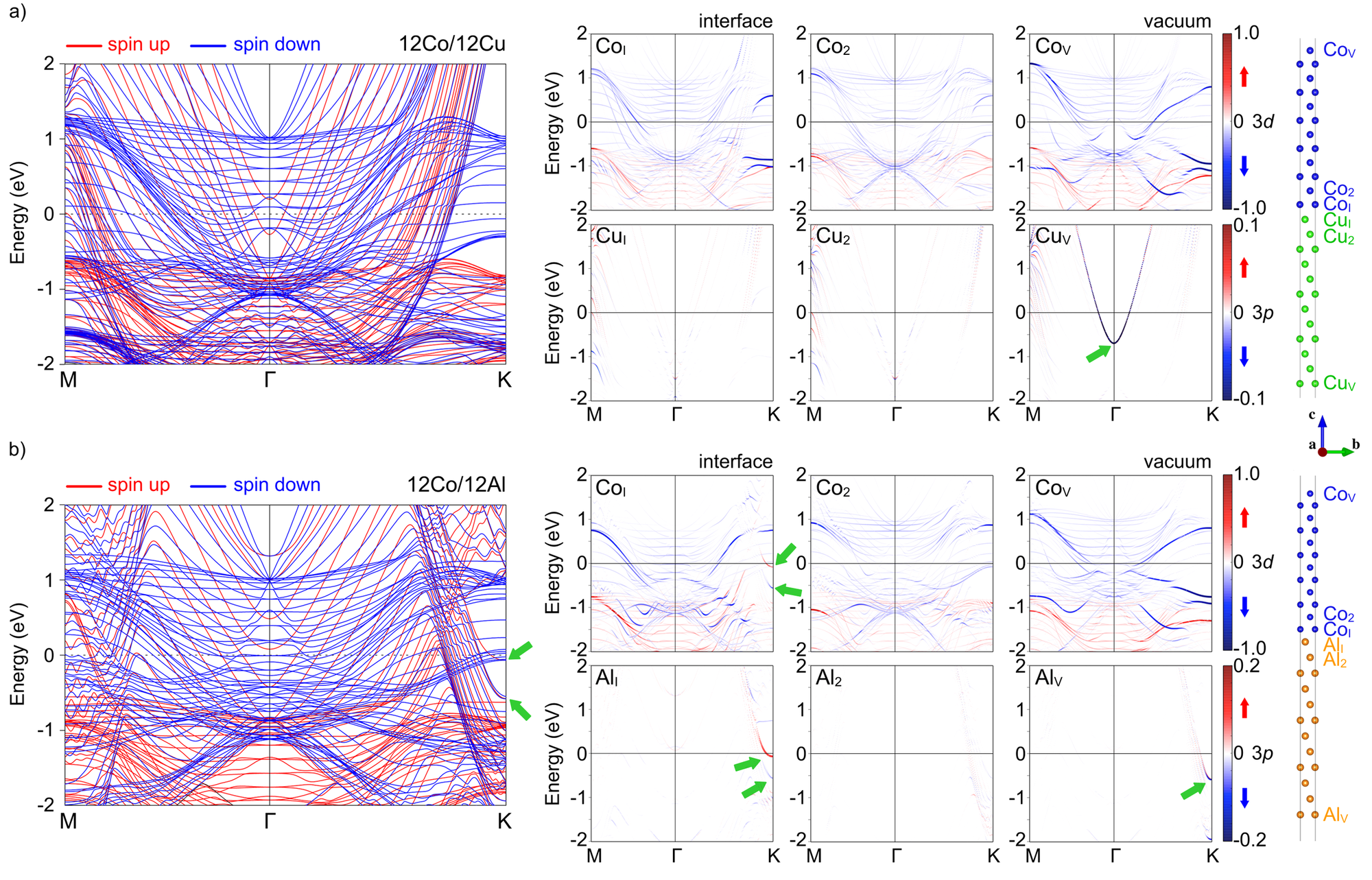}
\caption{Electronic band structures of a) 12Co(0001)/12Cu(111) and b) 12Co(0001)/12Al(111) heterostructures as obtained from GGA calculations without spin-orbit coupling, and layer-resolved projections of the band structure with spin-orbit coupling onto the corresponding atomic orbitals (Cu $3p$, Al $3p$, and Co $3d$ orbitals). Green arrows denote the surface states at the interfacial (I) and boundary (V) layers. Red and blue colorcodes stand for the spin-up and spin-down projections, respectively.}
\label{pic_1}
\end{figure}

\begin{figure}
\includegraphics[width=0.5\textwidth]{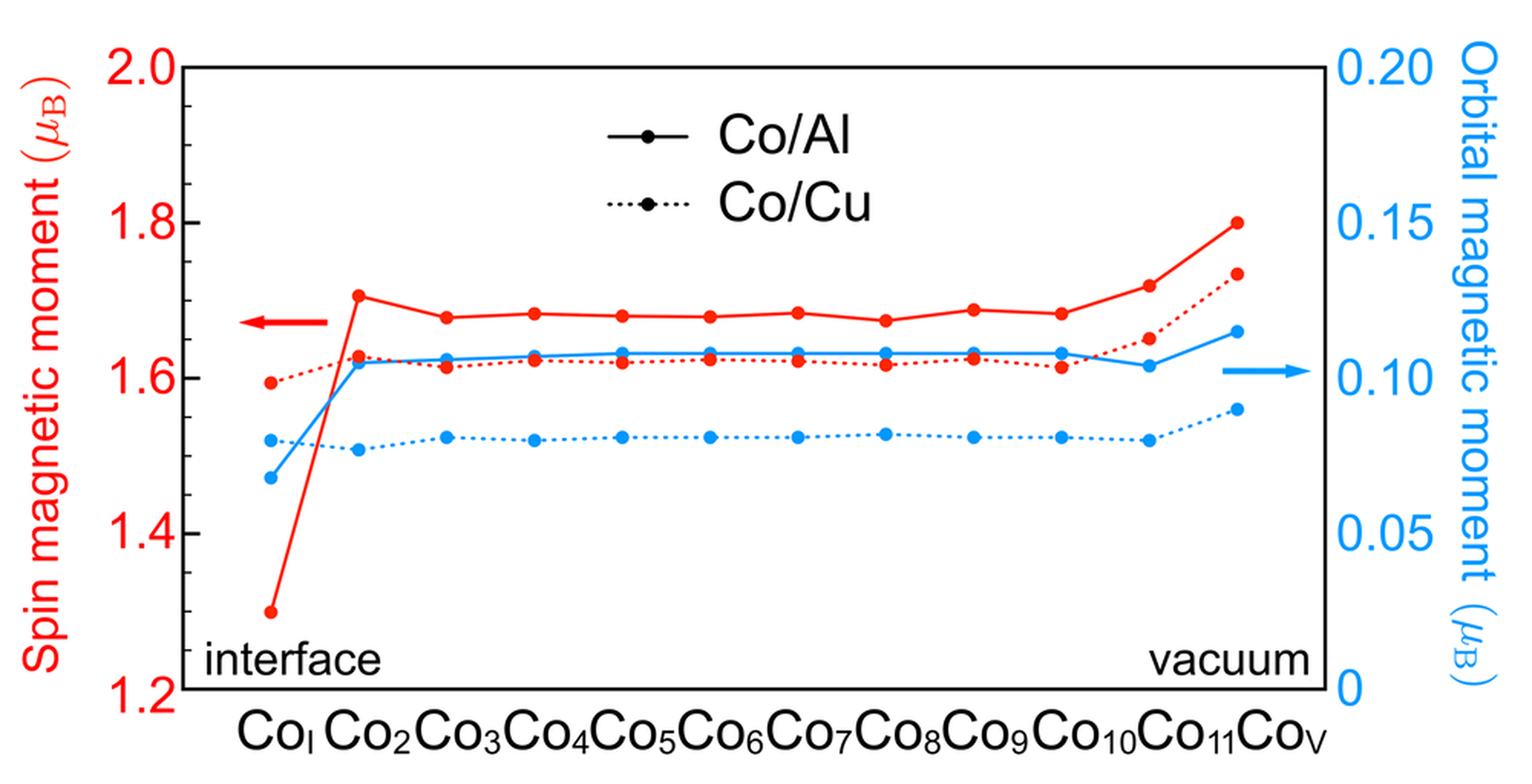}
\caption{Spin and orbital magnetic moments at Co layers in 12Co(0001)/12Cu(111) and 12Co(0001)/12Al(111) heterostructures as obtained from GGA calculations with spin-orbit coupling. Only $z$ components are shown, $x$ and $y$ components are zero at each layer.}
\label{pic_2}
\end{figure}

 The calculated electronic structures of Co/Cu and Co/Al heterostructures shown in Fig.~\ref{pic_1} suggest that the interfacial Co layers (Co$_{\mathrm{I}}$) possess several distinct features which are determined by hybridization with the adjacent nonmagnetic layers. In order to understand the essential differences in both interfaces, we can start by analyzing how the interfacial electronic states change in comparison to those of the off-lying layers. The spin non-polarized surface states are formed at the outmost Cu layer next to the vacuum region (Cu$_{\mathrm{V}}$ in Fig.~\ref{pic_1}a) which are located at the $\Gamma$ point of the Brillouin zone and have a predominantly Cu $p_{z}$ orbital character (see Figs. S2 and S3 in Supplementary Material). Similarly, the surface states are realized at the boundary Al layer (Al$_{\mathrm{V}}$ in Fig.~\ref{pic_1}b) which are, in contrast, located at the $K$ point and formed by the Al $p_{x}$ and $p_{y}$ orbitals (see Figs. S4 and S5 in Supplementary Material). At the interface, the hybridization primarily occurs between the Al/Cu $p$ and Co $d$ orbitals, being more pronounced in the vicinity of the $K$ point at the Co/Al interface due to the presence of the surface states. Indeed, while the hybridization clearly modifies the states at the interfacial layers in both systems,  one can see that the difference between the interfacial and boundary Co $3d$ states (Co$_{\mathrm{I}}$ and Co$_{\mathrm{V}}$) is more striking in the case of Co/Al heterostructure. Namely, the surface states become spin-polarized at the Co/Al interface due to the proximity effect (Al$_{\mathrm{I}}$ and Co$_{\mathrm{I}}$ in Fig.~\ref{pic_1}b), but vanish at the interfacial Cu layer (Cu$_{\mathrm{I}}$ in Fig.~\ref{pic_1}a). Moreover, as a result of the interfacial hybridization, both spin and orbital magnetic moments of Co presented in Fig.~\ref{pic_2} are relatively decreased at the Co/Al interface, as opposed to the Co/Cu interface, where neither of the magnetic moments reveals noticeable changes across the heterostructure. 

\par
\begin{figure}
\includegraphics[width=1.0\textwidth]{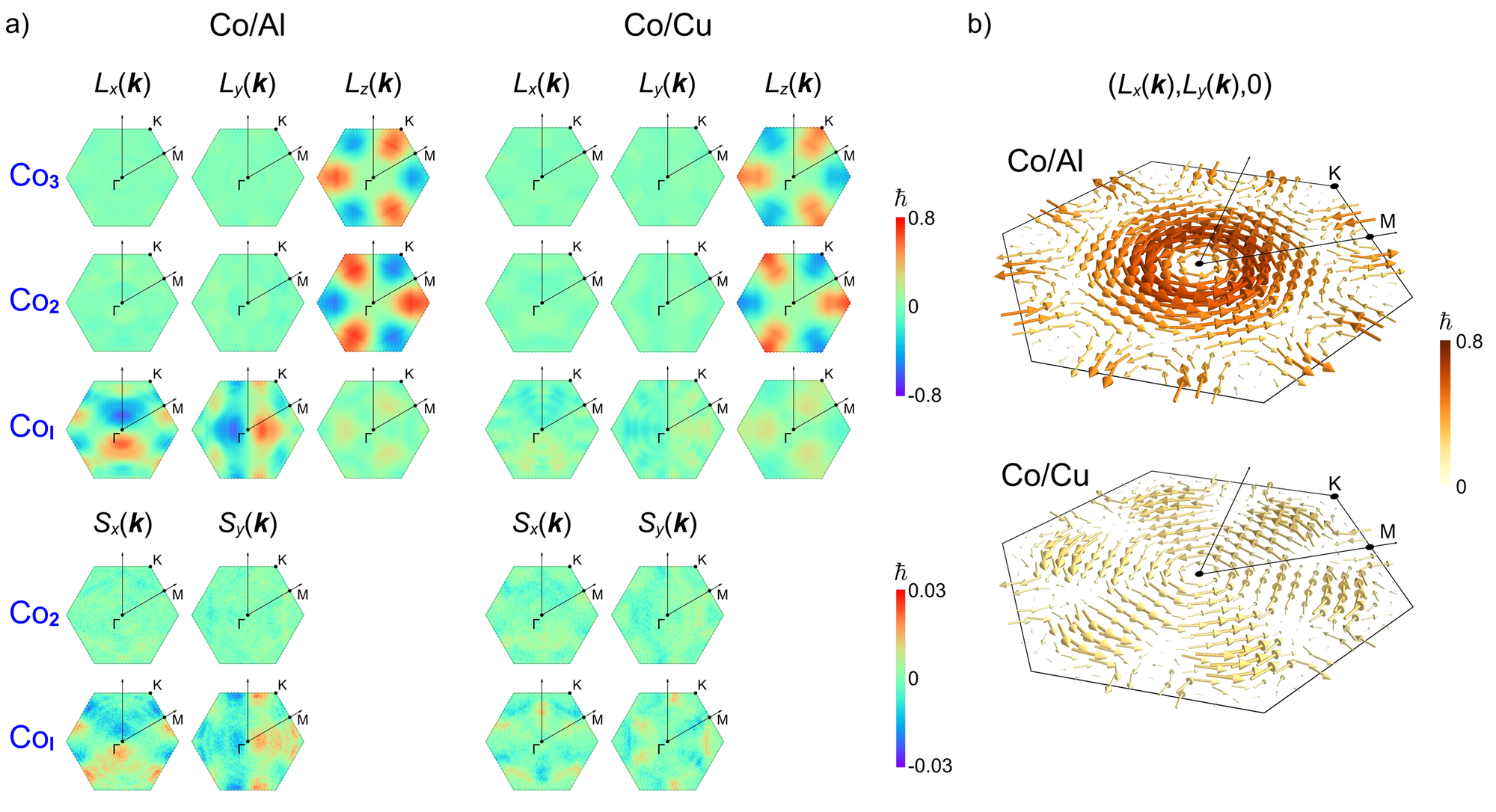}
\caption{a) Layer-resolved orbital and spin profiles in the Brillouin zone calculated for 12Co(0001)/12Al(111) and 12Co(0001)/12Cu(111) heterostructures with spin-orbit coupling. $\boldsymbol{L}(\boldsymbol{k})$ integrated over the Brillouin zone correspond to the orbital magnetic moments presented in Fig.~2. b) In-plane orbital textures at the interfacial Co layer. Arrows for the Co/Cu interface are magnified by two times for clarity. }
\label{pic_3}
\end{figure}

 Drastic changes in electronic and magnetic properties at the Co/Al and Co/Cu interfaces can be observed by calculating the orbital and spin profiles in reciprocal space, $\boldsymbol{L}(\boldsymbol{k})=\sum_{n}\boldsymbol{L}_{n}(\boldsymbol{k})f_{n\boldsymbol{k}}$ and $\boldsymbol{S}(\boldsymbol{k})=\sum_{n}\boldsymbol{S}_{n}(\boldsymbol{k})f_{n\boldsymbol{k}}$, respectively, where $f_{n\boldsymbol{k}}$ is the Fermi-Dirac occupation function with respect to the Fermi level, and $n$ runs over all bands. Here, $\boldsymbol{L}_{n}(\boldsymbol{k})=\langle u_{n\boldsymbol{k}}|\hat{\boldsymbol{L}}|u_{n\boldsymbol{k}}\rangle$ and $\boldsymbol{S}_{n}(\boldsymbol{k})=\langle u_{n\boldsymbol{k}}|\hat{\boldsymbol{S}}|u_{n\boldsymbol{k}}\rangle$ are the matrix elements of the orbital and spin angular momentum operators calculated by projecting the Bloch wavefunctions $|u_{n\boldsymbol{k}}\rangle=\sum_{\mu}^{\sigma}a^{\sigma}_{\mu n\boldsymbol{k}}|\mu\sigma\rangle$ onto atomic orbitals $|\mu\sigma\rangle$, with $\mu$ and $\sigma$ encoding orbital and spin indices, respectively. From the profiles shown in Fig.~\ref{pic_3}, one can see that both $\boldsymbol{S}(\boldsymbol{k})$ and $\boldsymbol{L}(\boldsymbol{k})$ exhibit a threefold rotational texture, in agreement with the $C_{3v}$ symmetry of the constructed heterostructures. The alternating pattern of $L_{z}(\boldsymbol{k})$ coming from non-zero matrix elements $\langle d_{yz}|\hat{L}_{z}|d_{zx}\rangle=i$ and $\langle d_{xy}|\hat{L}_{z}|d_{x^{2}-y^{2}}\rangle=2i$ is found to be strongly suppressed at the interfaces as compared to the interior layers. With the changes being seemingly identical at both interfaces, the average value of $L_{z}(\boldsymbol{k})$ over the Brillouin zone yields a decreased orbital moment for the Co/Al interface, in agreement with Fig.~\ref{pic_2}. More strikingly, the hybridization at the Co/Al interface turns out to give rise to an exceptionally large orbital texture with an in-plane helical locking at the interfacial Co layer, as opposed to the Co/Cu interface where the in-plane orbital texture is much less pronounced. Following the orbital character of the states hybridized at the interface, the largest contribution to the chiral orbital texture is found to arise from non-zero matrix elements $\langle d_{z^{2}}|\hat{L}_{x}|d_{yz}\rangle=\sqrt{3}i$ and $\langle d_{z^{2}}|\hat{L}_{y}|d_{zx}\rangle=-\sqrt{3}i$. Given that the hybridization occurs between the Al $p_{x}$ and $p_{y}$ and Co $d_{z^{2}}$, $d_{zx}$ and $d_{yz}$ orbitals, the emergence of the chiral orbital texture at the interfacial Co layer can be attributed to the so-called orbital Rashba effect\cite{PhysRevLett.107.156803,PhysRevB.85.195401,PhysRevB.85.195402,PhysRevB.87.041301,PhysRevB.103.L121113} as a result of peculiar hybridization with the surface states at the interface. It is worth noting that the chiral orbital texture at the Co/Al interface does not change qualitatively upon including the on-site Coulomb repulsion in the Co $3d$ orbitals, which changes the splitting between the fully occupied spin-up and partially occupied spin-down Co $3d$ states (see Figs.~S6--S13 with the results of GGA$+U$ calculations in Supplementary Material).\cite{PhysRevB.52.R5467,PhysRevB.57.1505,PhysRevB.76.035107,Cadi-Essadek2021} Neither does it change upon considering different stackings at the interface or changing the number of layers in the heterostructure (see Fig.~S14 in Supplementary Material), thus implying that the emergence of the chiral orbital texture can be regarded as a robust feature of the Co/Al interface due to the formaiton of the surface states. Finally, we note that the transverse spin profile $S_{x}(\boldsymbol{k})$ and $S_{y}(\boldsymbol{k})$ is found to have a similar chiral pattern at the interface of Co/Al heterostructure but is much smaller in magnitude (see also Fig.~S13 in Supplementary Material).

\par As follows from Fig.~\ref{pic_4}a, the in-plane orbital texture at the Co/Al interface is preserved without SOC, as opposed to the transverse spin texture which is identically zero in the absence of relativistic effects. Without SOC, $\boldsymbol{L}(\boldsymbol{k})=-\boldsymbol{L}(-\boldsymbol{k})$ holds true owing to time-reversal symmetry yielding the net orbital moment to be identically zero, when integrated over the Brillouin zone. In the presence of SOC, the magnetization direction is fixed along the easy axis, and time-reversal symmetry is broken leading to $\boldsymbol{L}(\boldsymbol{k})\ne-\boldsymbol{L}(-\boldsymbol{k})$ with a non-zero net orbital moment. To gain further insight into the origin of the chiral orbital texture, it is expedient to decompose the orbital moment as a sum of two contributions $\boldsymbol{L}(\boldsymbol{k})=\boldsymbol{L}^{(0)}(\boldsymbol{k})+\boldsymbol{L}^{\mathrm{(1)}}(\boldsymbol{k})$, where the first term $\boldsymbol{L}^{\mathrm{(0)}}(\boldsymbol{k})=\frac{1}{2}(\boldsymbol{L}(\boldsymbol{k})-\boldsymbol{L}(-\boldsymbol{k}))$ represents the time-reversal symmetric contribution, while the second one $\boldsymbol{L}^{\mathrm{(1)}}(\boldsymbol{k})=\frac{1}{2}(\boldsymbol{L}(\boldsymbol{k})+\boldsymbol{L}(-\boldsymbol{k}))$ can be identified with a nonvanishing net orbital moment with respect to the magnetization in the presence of SOC. Thus, $\boldsymbol{L}(\boldsymbol{k})=\boldsymbol{L}^{(0)}(\boldsymbol{k})$ and $\boldsymbol{L}^{\mathrm{(1)}}(\boldsymbol{k})=0$ in the absence of SOC, and SOC can generally contribute to both terms. From a practical point of view, one can find $\boldsymbol{L}^{\mathrm{(1)}}(\boldsymbol{k})$ either directly from fully relativistic electronic structure calculations including SOC using the definition above or by treating SOC within perturbation theory starting from the electronic structure calculated without SOC.\cite{PhysRevB.39.865,PhysRevB.47.14932,GerritvanderLaan1998} For the latter, $\boldsymbol{L}^{\mathrm{(1)}}(\boldsymbol{k})$ can be derived by decomposing the Bloch wavefunctions over the basis of cubic harmonics that yields to first order in SOC:
\begin{equation}
\boldsymbol{L}^{(1)}(\boldsymbol{k})=-\xi\,\mathrm{Re}\sum_{\nu\mu\mu'\nu'} \langle \nu|\hat{\boldsymbol{L}}|\nu'\rangle\langle\mu'|\hat{\boldsymbol{L}}\cdot\boldsymbol{m}|\mu\rangle (A_{\nu\mu\mu'\nu'}^{\uparrow\uparrow}(\boldsymbol{k})-A_{\nu\mu\mu'\nu'}^{\downarrow\downarrow}(\boldsymbol{k})),
\end{equation}
\noindent where $\xi=69.4$ meV is the SOC constant adopted for Co atoms,\cite{Dunn1961Jan,Cole1970Aug,PhysRevB.105.104427} and $A_{\nu\mu\mu'\nu'}^{\sigma\sigma'}(\boldsymbol{k})$ is the generalized susceptibility of linear response theory with $\mu$, $\mu'$, $\nu$, $\nu'$ and $\sigma$, $\sigma'$ being orbital and spin indices, respectively (see Section II in Supplementary Material). As one can see in Fig.~\ref{pic_4}b for the out-of-plane direction of magnetization $\boldsymbol{m}\parallel\boldsymbol{e}_{z}$, the inclusion of SOC results in a symmetric pancake-like profile for $L^{(1)}_{z}(\boldsymbol{k})$ giving the net orbital moment along the magnetization. In addition, the in-plane profile of $\boldsymbol{L}^{(1)}(\boldsymbol{k})$ shown in Fig.~\ref{pic_4}c has a higher-order winding of the orbital moment obeying the $C_{3v}$ symmetry and can be regarded as the SOC driven contribution to the orbital Rashba effect, similar to the spin Rashba effect.\cite{PhysRevB.85.075404} Importantly, the in-plane texture $\boldsymbol{L}^{(0)}(\boldsymbol{k})$ reveals only minor changes upon including SOC, and $\boldsymbol{L}^{(1)}(\boldsymbol{k})$ arising from SOC is found to be more than an order of magnitude smaller than $\boldsymbol{L}^{(0)}(\boldsymbol{k})$, thus implying that the emergence of the chiral orbital texture is essentially determined by the hybridization at the interface. 

\par
\begin{figure}
\includegraphics[width=0.5\textwidth]{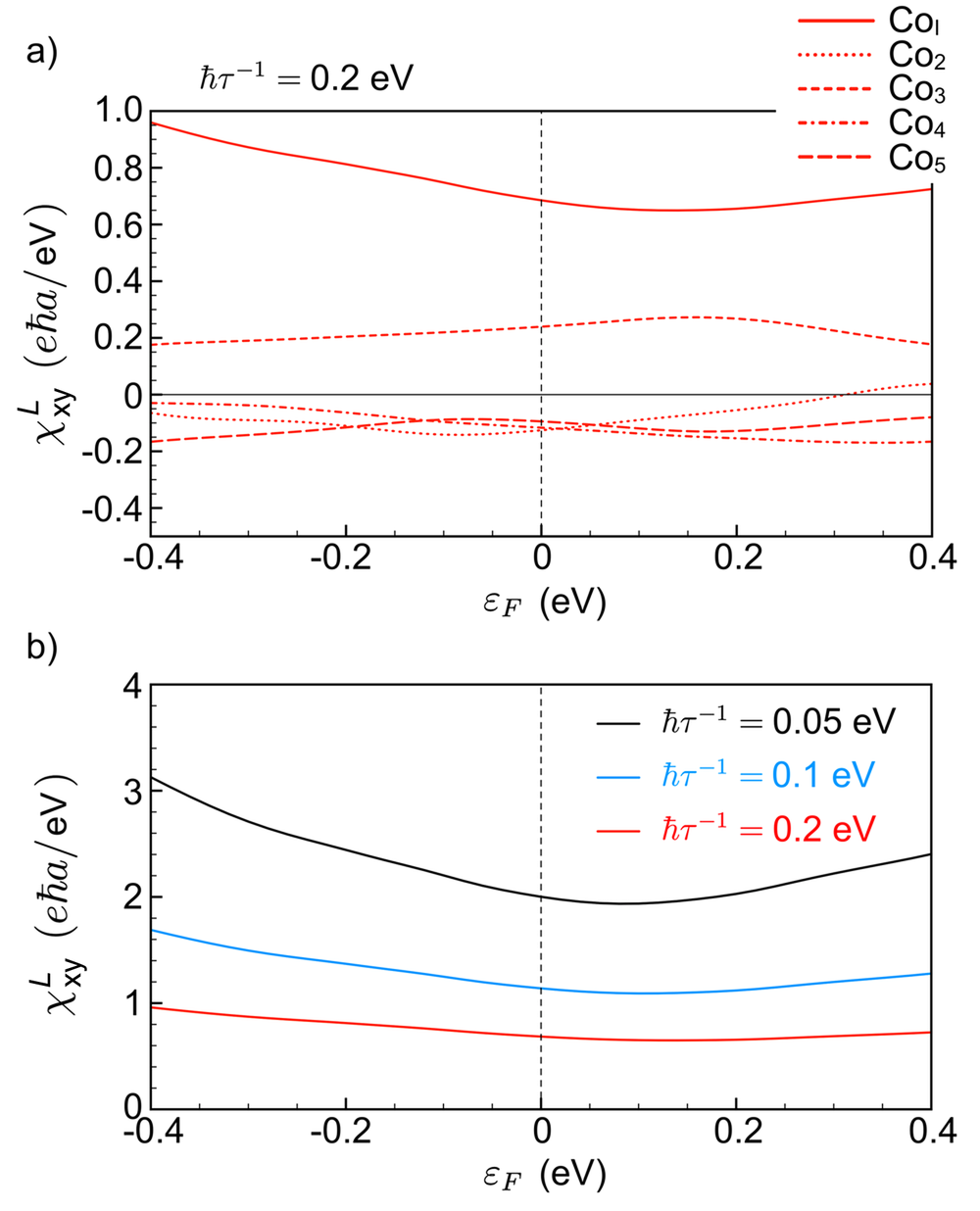}
\caption{a) Off-diagonal orbital magnetoelectric susceptibility $\chi_{xy}^{L}$ calculated for 8Co(0001)/6Al(111) heterostructure without spin-orbit coupling as a function of the Fermi level $\varepsilon_{F}$. b) $\chi_{yx}^{L}$ at the interfacial Co layer calculated for different relaxation times. Here, $a$ is the optimized lattice constant ($\sim$2.63~\AA~for 8Co(0001)/6Al(111) heterostructure). Calculations are performed on a 100$\times$100$\times$1 $k$-point mesh with the temperature factor $k_{\mathrm{B}}T=0.1$~eV.}
\label{pic_5}
\end{figure}

 The results of electronic structure calculations demonstrate that the Co/Al interface features a large in-plane chiral orbital texture. While both $L_{x}(\boldsymbol{k})$ and $L_{y}(\boldsymbol{k})$ cancel out in equilibrium for $\boldsymbol{m}\parallel\boldsymbol{e}_{z}$, an applied in-plane electric field can induce a finite orbital angular momentum, giving rise to the orbital Edelstein effect.\cite{Yoda2015Jul,Yoda2018Feb} The current induced orbital accumulation $\delta \boldsymbol{L}$ can be determined as a linear response to an electric field $\boldsymbol{E}$ by using Kubo theory as $\delta L_{\alpha}=\chi_{\alpha\beta}^{L}E_{\beta}$, where summation over repeated indices is implied ($\alpha$, $\beta=$ $x$, $y$, $z$). Starting from the one-particle picture $\hat{\mathcal{H}}_{\boldsymbol{k}}u_{n\boldsymbol{k}}(\boldsymbol{r})=\varepsilon_{n\boldsymbol{k}}u_{n\boldsymbol{k}}(\boldsymbol{r})$, where $\hat{\mathcal{H}}_{\boldsymbol{k}}$ is the Kohn-Sham Hamiltonian, and $u_{n\boldsymbol{k}}(\boldsymbol{r})$ is the Bloch wavefunction for band $n$ at wavevector $\boldsymbol{k}$ and energy $\varepsilon_{n\boldsymbol{k}}$, the rank-2 orbital magnetoelectric susceptibility tensor $\chi_{\alpha\beta}^{L}$ can be  written as a sum of two contributions:\cite{Salemi2019Nov,PhysRevMaterials.5.074407}
\begin{equation}
\delta L_{\alpha}^{\mathrm{intra}}=\chi_{\alpha\beta}^{L,\mathrm{intra}}E_{\beta}^{\phantom{L}}=-eE\tau\sum_{n\boldsymbol{k}}\frac{\partial f_{n\boldsymbol{k}}}{\partial\varepsilon_{n\boldsymbol{k}}}\langle u_{n\boldsymbol{k}}|\hat{L}_{\alpha}|u_{n\boldsymbol{k}}\rangle\langle u_{n\boldsymbol{k}}|\hat{v}_{\beta,\boldsymbol{k}}|u_{n\boldsymbol{k}}\rangle,
\end{equation}
\begin{equation}
\delta L_{\alpha}^{\mathrm{inter}}=\chi_{\alpha\beta}^{L,\mathrm{inter}}E_{\beta}^{\phantom{L}}=ie\hbar E\sum_{n\ne m,\boldsymbol{k}}\frac{(f_{n\boldsymbol{k}}-f_{m\boldsymbol{k}})\langle u_{m\boldsymbol{k}}|\hat{L}_{\alpha}|u_{n\boldsymbol{k}}\rangle\langle u_{n\boldsymbol{k}}|\hat{v}_{\beta,\boldsymbol{k}}|u_{m\boldsymbol{k}}\rangle}{(\varepsilon_{n\boldsymbol{k}}-\varepsilon_{m\boldsymbol{k}})(\varepsilon_{n\boldsymbol{k}}-\varepsilon_{m\boldsymbol{k}}-i\hbar\tau^{-1})},
\end{equation}
\noindent where $e$ $(<0)$ is the electron charge, $\hat{\boldsymbol{v}}_{\boldsymbol{k}}=\frac{1}{\hbar}\partial_{\boldsymbol{k}}\hat{\mathcal{H}}_{\boldsymbol{k}}$ is the velocity operator, and $\tau$ is the characteristic relaxation time defining the broadening of the electronic spectrum. The first and second terms correspond to the intraband contribution of the states at the Fermi surface and the interband transitions coming from the Fermi sea, respectively, summing up to $\chi_{\alpha\beta}^{L}=\chi_{\alpha\beta}^{L,\mathrm{intra}}+\chi_{\alpha\beta}^{L,\mathrm{inter}}$. In magnetic systems, the susceptibility tensor can be further decomposed into time-reversal-even and time-reversal-odd components as $\chi_{\alpha\beta}^{L}(\boldsymbol{m})=\chi_{\alpha\beta}^{\mathrm{even}}(\boldsymbol{m})+\chi_{\alpha\beta}^{\mathrm{odd}}(\boldsymbol{m})$, with $\chi_{\alpha\beta}^{\mathrm{even}}(\boldsymbol{m})=\frac{1}{2}(\chi_{\alpha\beta}^{L}(\boldsymbol{m})+\chi_{\alpha\beta}^{L}(-\boldsymbol{m}))$ and $\chi_{\alpha\beta}^{\mathrm{odd}}(\boldsymbol{m})=\frac{1}{2}(\chi_{\alpha\beta}^{L}(\boldsymbol{m})-\chi_{\alpha\beta}^{L}(-\boldsymbol{m}))$. Taking the out-of-plane direction of $\boldsymbol{m}$ and the symmetry of Co/Al heterostructures by construction as the $C_{3v}$ group, one can show that the only non-zero components of $\chi_{\alpha\beta}^{L}(\boldsymbol{m})$ are the off-diagonal $\chi_{xy}^{\mathrm{even}}(\boldsymbol{m})=-\chi_{yx}^{\mathrm{even}}(\boldsymbol{m})$ and diagonal $\chi_{xx}^{\mathrm{odd}}(\boldsymbol{m})=\chi_{yy}^{\mathrm{odd}}(\boldsymbol{m})\ne\chi_{zz}^{\mathrm{odd}}(\boldsymbol{m})$ susceptibilities (see Section~IIIa in Supplementary Material).\cite{PhysRev.142.318,PhysRevB.92.155138} Following the symmetry analysis for the in-plane orbital texture $\boldsymbol{L}(\boldsymbol{k})$, it is straightforward to note that $\boldsymbol{L}^{(0)}(\boldsymbol{k})$ and $\boldsymbol{L}^{(1)}(\boldsymbol{k})$ determine $\chi_{xy}^{\mathrm{even}}(\boldsymbol{m})$ and $\chi_{xx}^{\mathrm{odd}}(\boldsymbol{m})$, respectively. Since $\boldsymbol{L}^{(1)}(\boldsymbol{k})$ is much smaller than $\boldsymbol{L}^{(0)}(\boldsymbol{k})$, the off-diagonal susceptibility $\chi_{xy}^{\mathrm{even}}(\boldsymbol{m})$ originating from the chiral orbital texture $\boldsymbol{L}^{(0)}(\boldsymbol{k})$ can be considered to generate the leading non-equilibrium orbital response at the Co/Al interface (see Section IIID in Supplementary Material). Calculations of the orbital susceptibility tensor are performed using the Wannier interpolation\cite{PhysRevB.56.12847,RevModPhys.84.1419} of the electronic structure along the lines with previous studies on the anomalous and spin Hall susceptibilities (see Section IIIb in Supplementary Material).\cite{PhysRevB.74.195118,PhysRevB.99.235113,PhysRevB.98.214402} Given the smallness of $\boldsymbol{L}^{(1)}(\boldsymbol{k})$ and that $\boldsymbol{L}^{(0)}(\boldsymbol{k})$ does not depend on the magnetization to first order in SOC, $\chi_{xy}^{L}$ can be calculated without SOC. The results shown in Fig.~\ref{pic_5} illustrate that the orbital magnetoelectric susceptibility at the interfacial Co layer stands out in magnitude in comparison with the interior Co layers. The diagonal components $\chi_{xx}^{\mathrm{even}}$ are identicaly zero in agreement with the symmetry consideration.

\par The induced orbital accumulation $\delta \boldsymbol{L}$ couples to the spin accumulation $\delta \boldsymbol{S}$ by means of SOC and can exert a torque $\boldsymbol{\tau}=\boldsymbol{M}\times\delta\boldsymbol{B}_{\mathrm{xc}}$ on the magnetization, where $\delta\boldsymbol{B}_{\mathrm{xc}}\sim\delta\boldsymbol{S}$ is a change in the effective exchange-correlation magnetic field caused by the non-equilibrium spin accumulation. Given the in-plane electric field $\boldsymbol{E}$ and the normal direction $\boldsymbol{e}_{z}$, the spin accumulation $\delta \boldsymbol{S}$ can be decomposed into the $E$-transverse $\sim\boldsymbol{E}\times\boldsymbol{e}_{z}$ and $M$-transverse $\sim \boldsymbol{m}\times(\boldsymbol{E}\times\boldsymbol{e}_{z})$ components, where the former is time-reversal even and does not depend on the magnetization direction, while the latter depends on the orientation of $\boldsymbol{m}$ with respect to $\boldsymbol{E}$ and $\boldsymbol{e}_{z}$. Thus, it follows that the field-like $\boldsymbol{\tau}_{\mathrm{FL}}\sim \boldsymbol{m}\times(\boldsymbol{E}\times\boldsymbol{e}_{z})$ and the damping-like $\boldsymbol{\tau}_{\mathrm{DL}}\sim \boldsymbol{m}\times(\boldsymbol{m}\times(\boldsymbol{E}\times\boldsymbol{e}_{z}))$ torques can be expressed through the off-diagonal $\chi_{xy}^{\mathrm{even}}(\boldsymbol{m})$ and diagonal $\chi_{xx}^{\mathrm{odd}}(\boldsymbol{m})$ components, respectively. Using the results of linear response theory, one can estimate the effective magnetic field producing $\boldsymbol{\tau}_{\mathrm{FL}}$ as $B_{\mathrm{FL}}=\xi \chi^{L}_{xy}E/(M_{s}d_{F})$ (see Section IIIc of Supplementary Materials) that for the thickness $d_{F}\sim10^{-9}$~m, $E\sim2.5\cdot10^{4}$~V/m (for the spin current density $J^{S}\sim10^{11}$~A/m$^{2}$ in the experimental results), saturation magnetization $M_{s}\sim1.4\cdot10^{6}$~A/m, $\xi\sim10^{-20}$~J, and $\chi^{L}_{xy}\sim2.0\cdot10^{10}\hbar$~m/V (for $\hbar\tau^{-1}=0.05$~eV and $\varepsilon_{\mathrm{F}}=0$) gives $B_{\mathrm{FL}}~3.6$~mT, of comparable magnitude with the increase of $B_{\mathrm{FL}}$ by about 7~mT with respect to its value around 0.6~mT before insertion of Al.\cite{acs.nanolett.2c05091} With the diagonal components $\chi_{\alpha\alpha}^{L}$ being an order of magnitude smaller, the associated contribution to the damping-like torque can be regarded small compared to  $\boldsymbol{\tau}_{\mathrm{FL}}$. That being said, $\boldsymbol{\tau}_{\mathrm{DL}}$ is also likely to originate either from the diffusion of the orbital accumulation into the interior layers, which is further transferred to the magnetization by SOC-induced orbital precession, or indirectly, by conversion to the spin accumulation and subsequent transfer to the magnetization by spin precession. This generation of $\boldsymbol{\tau}_{\mathrm{DL}}$ by out-of-equilibrium processes, not calculated here, is expected to be smaller than the field-like torque produced by SOC directly from the interfacial orbital accumulation, the predominance of which is seen in the experiments. Our analysis is also in good agreement with micromagnetic estimates suggesting that large spin accumulation induced by a Rashba-like interaction at the interface is required to fit the experimental data.\cite{acs.nanolett.2c05091}

\par In conclusion, the results of first-principles calculations demonstrate that the Co/Al interface features a large orbital texture with the in-plane helical locking of the orbital moment in reciprocal space, which is found to be much smaller at the Co/Cu interface. The origin of the chiral orbital texture is attributed to the orbital Rashba effect due to the formation of the surface states at the interface with smaller higher-order contributions coming from SOC. Our calculations unveil that the orbital texture is responsible for the conversion effects in response to an applied electric current that produce large field-like torques at the Co/Al interface, thus providing a rigorous theoretical interpretation for recent experiments\cite{acs.nanolett.2c05091} and advancing our understanding of transport phenomena in all-metallic systems with light elements for spinorbitronic based technologies.

\par \textbf{Supporting Information.} Computational details of electronic structure calculations; perturbation theory for the orbital magnetic moment; calculations of the orbital magnetoelectric tensor.

\par \textbf{Acknowledgements.} The authors thank Mar\'ia Blanco-Rey and Andres Arnau for stimulating discussions. This study has been supported by the French National Research Agency under the project ``ORION" ANR-20-CE30-0022-02, by a France 2030 government grant managed by the French National Research Agency PEPR SPIN ANR-22-EXSP 0009 (SPINTHEORY), by the European Horizon Europe Framework Programme under an EC Grant Agreement N°101129641 ``OBELIX", and by the Jean d'Alembert fellowship program from Universit\'e Paris-Saclay.

\bibliography{achemso-demo}

\providecommand{\latin}[1]{#1}
\makeatletter
\providecommand{\doi}
  {\begingroup\let\do\@makeother\dospecials
  \catcode`\{=1 \catcode`\}=2 \doi@aux}
\providecommand{\doi@aux}[1]{\endgroup\texttt{#1}}
\makeatother
\providecommand*\mcitethebibliography{\thebibliography}
\csname @ifundefined\endcsname{endmcitethebibliography}
  {\let\endmcitethebibliography\endthebibliography}{}
\begin{mcitethebibliography}{62}
\providecommand*\natexlab[1]{#1}
\providecommand*\mciteSetBstSublistMode[1]{}
\providecommand*\mciteSetBstMaxWidthForm[2]{}
\providecommand*\mciteBstWouldAddEndPuncttrue
  {\def\EndOfBibitem{\unskip.}}
\providecommand*\mciteBstWouldAddEndPunctfalse
  {\let\EndOfBibitem\relax}
\providecommand*\mciteSetBstMidEndSepPunct[3]{}
\providecommand*\mciteSetBstSublistLabelBeginEnd[3]{}
\providecommand*\EndOfBibitem{}
\mciteSetBstSublistMode{f}
\mciteSetBstMaxWidthForm{subitem}{(\alph{mcitesubitemcount})}
\mciteSetBstSublistLabelBeginEnd
  {\mcitemaxwidthsubitemform\space}
  {\relax}
  {\relax}

\bibitem[Manchon \latin{et~al.}(2019)Manchon, \ifmmode~\check{Z}\else
  \v{Z}\fi{}elezn\'y, Miron, Jungwirth, Sinova, Thiaville, Garello, and
  Gambardella]{RevModPhys.91.035004}
Manchon,~A.; \ifmmode~\check{Z}\else \v{Z}\fi{}elezn\'y,~J.; Miron,~I.~M.;
  Jungwirth,~T.; Sinova,~J.; Thiaville,~A.; Garello,~K.; Gambardella,~P.
  Current-induced spin-orbit torques in ferromagnetic and antiferromagnetic
  systems. \emph{Rev. Mod. Phys.} \textbf{2019}, \emph{91}, 035004\relax
\mciteBstWouldAddEndPuncttrue
\mciteSetBstMidEndSepPunct{\mcitedefaultmidpunct}
{\mcitedefaultendpunct}{\mcitedefaultseppunct}\relax
\EndOfBibitem
\bibitem[{D'Yakonov} and {Perel'}(1971){D'Yakonov}, and
  {Perel'}]{1971ZhPmR..13..657D}
{D'Yakonov},~M.~I.; {Perel'},~V.~I. {Possibility of Orienting Electron Spins
  with Current}. \emph{ZhETF Pisma Redaktsiiu} \textbf{1971}, \emph{13},
  657\relax
\mciteBstWouldAddEndPuncttrue
\mciteSetBstMidEndSepPunct{\mcitedefaultmidpunct}
{\mcitedefaultendpunct}{\mcitedefaultseppunct}\relax
\EndOfBibitem
\bibitem[Hirsch(1999)]{PhysRevLett.83.1834}
Hirsch,~J.~E. Spin Hall Effect. \emph{Phys. Rev. Lett.} \textbf{1999},
  \emph{83}, 1834--1837\relax
\mciteBstWouldAddEndPuncttrue
\mciteSetBstMidEndSepPunct{\mcitedefaultmidpunct}
{\mcitedefaultendpunct}{\mcitedefaultseppunct}\relax
\EndOfBibitem
\bibitem[Wunderlich \latin{et~al.}(2005)Wunderlich, Kaestner, Sinova, and
  Jungwirth]{PhysRevLett.94.047204}
Wunderlich,~J.; Kaestner,~B.; Sinova,~J.; Jungwirth,~T. Experimental
  Observation of the Spin-Hall Effect in a Two-Dimensional Spin-Orbit Coupled
  Semiconductor System. \emph{Phys. Rev. Lett.} \textbf{2005}, \emph{94},
  047204\relax
\mciteBstWouldAddEndPuncttrue
\mciteSetBstMidEndSepPunct{\mcitedefaultmidpunct}
{\mcitedefaultendpunct}{\mcitedefaultseppunct}\relax
\EndOfBibitem
\bibitem[Sinova \latin{et~al.}(2015)Sinova, Valenzuela, Wunderlich, Back, and
  Jungwirth]{RevModPhys.87.1213}
Sinova,~J.; Valenzuela,~S.~O.; Wunderlich,~J.; Back,~C.~H.; Jungwirth,~T. Spin
  Hall effects. \emph{Rev. Mod. Phys.} \textbf{2015}, \emph{87},
  1213--1260\relax
\mciteBstWouldAddEndPuncttrue
\mciteSetBstMidEndSepPunct{\mcitedefaultmidpunct}
{\mcitedefaultendpunct}{\mcitedefaultseppunct}\relax
\EndOfBibitem
\bibitem[Bychkov and Rashba(1984)Bychkov, and Rashba]{Bychkov1984Jan}
Bychkov,~{\relax Yu}.~A.; Rashba,~{\ifmmode\acute{E}\else\'{E}\fi}.~I.
  {Properties of a 2D electron gas with lifted spectral degeneracy}.
  \emph{Soviet Journal of Experimental and Theoretical Physics Letters}
  \textbf{1984}, \emph{39}, 78\relax
\mciteBstWouldAddEndPuncttrue
\mciteSetBstMidEndSepPunct{\mcitedefaultmidpunct}
{\mcitedefaultendpunct}{\mcitedefaultseppunct}\relax
\EndOfBibitem
\bibitem[Edelstein(1990)]{EDELSTEIN1990233}
Edelstein,~V. Spin polarization of conduction electrons induced by electric
  current in two-dimensional asymmetric electron systems. \emph{Solid State
  Communications} \textbf{1990}, \emph{73}, 233--235\relax
\mciteBstWouldAddEndPuncttrue
\mciteSetBstMidEndSepPunct{\mcitedefaultmidpunct}
{\mcitedefaultendpunct}{\mcitedefaultseppunct}\relax
\EndOfBibitem
\bibitem[Garello \latin{et~al.}(2013)Garello, Miron, Avci, Freimuth, Mokrousov,
  Bl{\ifmmode\ddot{u}\else\"{u}\fi}gel, Auffret, Boulle, Gaudin, and
  Gambardella]{Garello2013Aug}
Garello,~K.; Miron,~I.~M.; Avci,~C.~O.; Freimuth,~F.; Mokrousov,~Y.;
  Bl{\ifmmode\ddot{u}\else\"{u}\fi}gel,~S.; Auffret,~S.; Boulle,~O.;
  Gaudin,~G.; Gambardella,~P. {Symmetry and magnitude of spin-orbit torques in
  ferromagnetic heterostructures}. \emph{Nat. Nanotechnol.} \textbf{2013},
  \emph{8}, 587--593\relax
\mciteBstWouldAddEndPuncttrue
\mciteSetBstMidEndSepPunct{\mcitedefaultmidpunct}
{\mcitedefaultendpunct}{\mcitedefaultseppunct}\relax
\EndOfBibitem
\bibitem[Liu \latin{et~al.}(2012)Liu, Pai, Li, Tseng, Ralph, and
  Buhrman]{Liu2012May}
Liu,~L.; Pai,~C.-F.; Li,~Y.; Tseng,~H.~W.; Ralph,~D.~C.; Buhrman,~R.~A.
  {Spin-Torque Switching with the Giant Spin Hall Effect of Tantalum}.
  \emph{Science} \textbf{2012}, \emph{336}, 555--558\relax
\mciteBstWouldAddEndPuncttrue
\mciteSetBstMidEndSepPunct{\mcitedefaultmidpunct}
{\mcitedefaultendpunct}{\mcitedefaultseppunct}\relax
\EndOfBibitem
\bibitem[Fan \latin{et~al.}(2014)Fan, Celik, Wu, Ni, Lee, Lorenz, and
  Xiao]{Fan2014Jan}
Fan,~X.; Celik,~H.; Wu,~J.; Ni,~C.; Lee,~K.-J.; Lorenz,~V.~O.; Xiao,~J.~Q.
  {Quantifying interface and bulk contributions to spin{\textendash}orbit
  torque in magnetic bilayers}. \emph{Nat. Commun.} \textbf{2014}, \emph{5},
  1--8\relax
\mciteBstWouldAddEndPuncttrue
\mciteSetBstMidEndSepPunct{\mcitedefaultmidpunct}
{\mcitedefaultendpunct}{\mcitedefaultseppunct}\relax
\EndOfBibitem
\bibitem[Mihai~Miron \latin{et~al.}(2010)Mihai~Miron, Gaudin, Auffret, Rodmacq,
  Schuhl, Pizzini, Vogel, and Gambardella]{MihaiMiron2010Mar}
Mihai~Miron,~I.; Gaudin,~G.; Auffret,~S.; Rodmacq,~B.; Schuhl,~A.; Pizzini,~S.;
  Vogel,~J.; Gambardella,~P. {Current-driven spin torque induced by the Rashba
  effect in a ferromagnetic metal layer}. \emph{Nat. Mater.} \textbf{2010},
  \emph{9}, 230--234\relax
\mciteBstWouldAddEndPuncttrue
\mciteSetBstMidEndSepPunct{\mcitedefaultmidpunct}
{\mcitedefaultendpunct}{\mcitedefaultseppunct}\relax
\EndOfBibitem
\bibitem[Miron \latin{et~al.}(2011)Miron, Garello, Gaudin, Zermatten, Costache,
  Auffret, Bandiera, Rodmacq, Schuhl, and Gambardella]{Miron2011Aug}
Miron,~I.~M.; Garello,~K.; Gaudin,~G.; Zermatten,~P.-J.; Costache,~M.~V.;
  Auffret,~S.; Bandiera,~S.; Rodmacq,~B.; Schuhl,~A.; Gambardella,~P.
  {Perpendicular switching of a single ferromagnetic layer induced by in-plane
  current injection}. \emph{Nature} \textbf{2011}, \emph{476}, 189--193\relax
\mciteBstWouldAddEndPuncttrue
\mciteSetBstMidEndSepPunct{\mcitedefaultmidpunct}
{\mcitedefaultendpunct}{\mcitedefaultseppunct}\relax
\EndOfBibitem
\bibitem[Liu \latin{et~al.}(2011)Liu, Moriyama, Ralph, and
  Buhrman]{PhysRevLett.106.036601}
Liu,~L.; Moriyama,~T.; Ralph,~D.~C.; Buhrman,~R.~A. Spin-Torque Ferromagnetic
  Resonance Induced by the Spin Hall Effect. \emph{Phys. Rev. Lett.}
  \textbf{2011}, \emph{106}, 036601\relax
\mciteBstWouldAddEndPuncttrue
\mciteSetBstMidEndSepPunct{\mcitedefaultmidpunct}
{\mcitedefaultendpunct}{\mcitedefaultseppunct}\relax
\EndOfBibitem
\bibitem[Liu \latin{et~al.}(2012)Liu, Lee, Gudmundsen, Ralph, and
  Buhrman]{PhysRevLett.109.096602}
Liu,~L.; Lee,~O.~J.; Gudmundsen,~T.~J.; Ralph,~D.~C.; Buhrman,~R.~A.
  Current-Induced Switching of Perpendicularly Magnetized Magnetic Layers Using
  Spin Torque from the Spin Hall Effect. \emph{Phys. Rev. Lett.} \textbf{2012},
  \emph{109}, 096602\relax
\mciteBstWouldAddEndPuncttrue
\mciteSetBstMidEndSepPunct{\mcitedefaultmidpunct}
{\mcitedefaultendpunct}{\mcitedefaultseppunct}\relax
\EndOfBibitem
\bibitem[Miron \latin{et~al.}(2011)Miron, Moore, Szambolics, Buda-Prejbeanu,
  Auffret, Rodmacq, Pizzini, Vogel, Bonfim, Schuhl, and Gaudin]{Miron2011Jun}
Miron,~I.~M.; Moore,~T.; Szambolics,~H.; Buda-Prejbeanu,~L.~D.; Auffret,~S.;
  Rodmacq,~B.; Pizzini,~S.; Vogel,~J.; Bonfim,~M.; Schuhl,~A.; Gaudin,~G. {Fast
  current-induced domain-wall motion controlled by the Rashba effect}.
  \emph{Nat. Mater.} \textbf{2011}, \emph{10}, 419--423\relax
\mciteBstWouldAddEndPuncttrue
\mciteSetBstMidEndSepPunct{\mcitedefaultmidpunct}
{\mcitedefaultendpunct}{\mcitedefaultseppunct}\relax
\EndOfBibitem
\bibitem[Thiaville \latin{et~al.}(2012)Thiaville, Rohart,
  Ju{\ifmmode\acute{e}\else\'{e}\fi}, Cros, and Fert]{Thiaville2012Dec}
Thiaville,~A.; Rohart,~S.;
  Ju{\ifmmode\acute{e}\else\'{e}\fi},~{\ifmmode\acute{E}\else\'{E}\fi}.;
  Cros,~V.; Fert,~A. {Dynamics of Dzyaloshinskii domain walls in ultrathin
  magnetic films}. \emph{Europhys. Lett.} \textbf{2012}, \emph{100},
  57002\relax
\mciteBstWouldAddEndPuncttrue
\mciteSetBstMidEndSepPunct{\mcitedefaultmidpunct}
{\mcitedefaultendpunct}{\mcitedefaultseppunct}\relax
\EndOfBibitem
\bibitem[Ryu \latin{et~al.}(2013)Ryu, Thomas, Yang, and Parkin]{Ryu2013Jul}
Ryu,~K.-S.; Thomas,~L.; Yang,~S.-H.; Parkin,~S. {Chiral spin torque at magnetic
  domain walls}. \emph{Nat. Nanotechnol.} \textbf{2013}, \emph{8},
  527--533\relax
\mciteBstWouldAddEndPuncttrue
\mciteSetBstMidEndSepPunct{\mcitedefaultmidpunct}
{\mcitedefaultendpunct}{\mcitedefaultseppunct}\relax
\EndOfBibitem
\bibitem[Khvalkovskiy \latin{et~al.}(2013)Khvalkovskiy, Cros, Apalkov, Nikitin,
  Krounbi, Zvezdin, Anane, Grollier, and Fert]{PhysRevB.87.020402}
Khvalkovskiy,~A.~V.; Cros,~V.; Apalkov,~D.; Nikitin,~V.; Krounbi,~M.;
  Zvezdin,~K.~A.; Anane,~A.; Grollier,~J.; Fert,~A. Matching domain-wall
  configuration and spin-orbit torques for efficient domain-wall motion.
  \emph{Phys. Rev. B} \textbf{2013}, \emph{87}, 020402\relax
\mciteBstWouldAddEndPuncttrue
\mciteSetBstMidEndSepPunct{\mcitedefaultmidpunct}
{\mcitedefaultendpunct}{\mcitedefaultseppunct}\relax
\EndOfBibitem
\bibitem[Tanaka \latin{et~al.}(2008)Tanaka, Kontani, Naito, Naito, Hirashima,
  Yamada, and Inoue]{PhysRevB.77.165117}
Tanaka,~T.; Kontani,~H.; Naito,~M.; Naito,~T.; Hirashima,~D.~S.; Yamada,~K.;
  Inoue,~J. Intrinsic spin Hall effect and orbital Hall effect in $4d$ and $5d$
  transition metals. \emph{Phys. Rev. B} \textbf{2008}, \emph{77}, 165117\relax
\mciteBstWouldAddEndPuncttrue
\mciteSetBstMidEndSepPunct{\mcitedefaultmidpunct}
{\mcitedefaultendpunct}{\mcitedefaultseppunct}\relax
\EndOfBibitem
\bibitem[Kontani \latin{et~al.}(2008)Kontani, Tanaka, Hirashima, Yamada, and
  Inoue]{PhysRevLett.100.096601}
Kontani,~H.; Tanaka,~T.; Hirashima,~D.~S.; Yamada,~K.; Inoue,~J. Giant
  Intrinsic Spin and Orbital Hall Effects in
  ${\mathrm{Sr}}_{2}M{\mathrm{O}}_{4}$ ($M=\mathrm{Ru}$, Rh, Mo). \emph{Phys.
  Rev. Lett.} \textbf{2008}, \emph{100}, 096601\relax
\mciteBstWouldAddEndPuncttrue
\mciteSetBstMidEndSepPunct{\mcitedefaultmidpunct}
{\mcitedefaultendpunct}{\mcitedefaultseppunct}\relax
\EndOfBibitem
\bibitem[Park \latin{et~al.}(2011)Park, Kim, Yu, Han, and
  Kim]{PhysRevLett.107.156803}
Park,~S.~R.; Kim,~C.~H.; Yu,~J.; Han,~J.~H.; Kim,~C. Orbital-Angular-Momentum
  Based Origin of Rashba-Type Surface Band Splitting. \emph{Phys. Rev. Lett.}
  \textbf{2011}, \emph{107}, 156803\relax
\mciteBstWouldAddEndPuncttrue
\mciteSetBstMidEndSepPunct{\mcitedefaultmidpunct}
{\mcitedefaultendpunct}{\mcitedefaultseppunct}\relax
\EndOfBibitem
\bibitem[Park \latin{et~al.}(2012)Park, Kim, Rhim, and Han]{PhysRevB.85.195401}
Park,~J.-H.; Kim,~C.~H.; Rhim,~J.-W.; Han,~J.~H. Orbital Rashba effect and its
  detection by circular dichroism angle-resolved photoemission spectroscopy.
  \emph{Phys. Rev. B} \textbf{2012}, \emph{85}, 195401\relax
\mciteBstWouldAddEndPuncttrue
\mciteSetBstMidEndSepPunct{\mcitedefaultmidpunct}
{\mcitedefaultendpunct}{\mcitedefaultseppunct}\relax
\EndOfBibitem
\bibitem[Kim \latin{et~al.}(2012)Kim, Kim, Kim, Jung, Kim, Koh, Arita, Shimada,
  Namatame, Taniguchi, Yu, and Kim]{PhysRevB.85.195402}
Kim,~B.; Kim,~C.~H.; Kim,~P.; Jung,~W.; Kim,~Y.; Koh,~Y.; Arita,~M.;
  Shimada,~K.; Namatame,~H.; Taniguchi,~M.; Yu,~J.; Kim,~C. Spin and orbital
  angular momentum structure of Cu(111) and Au(111) surface states. \emph{Phys.
  Rev. B} \textbf{2012}, \emph{85}, 195402\relax
\mciteBstWouldAddEndPuncttrue
\mciteSetBstMidEndSepPunct{\mcitedefaultmidpunct}
{\mcitedefaultendpunct}{\mcitedefaultseppunct}\relax
\EndOfBibitem
\bibitem[Park \latin{et~al.}(2013)Park, Kim, Lee, and Han]{PhysRevB.87.041301}
Park,~J.-H.; Kim,~C.~H.; Lee,~H.-W.; Han,~J.~H. Orbital chirality and Rashba
  interaction in magnetic bands. \emph{Phys. Rev. B} \textbf{2013}, \emph{87},
  041301(R)\relax
\mciteBstWouldAddEndPuncttrue
\mciteSetBstMidEndSepPunct{\mcitedefaultmidpunct}
{\mcitedefaultendpunct}{\mcitedefaultseppunct}\relax
\EndOfBibitem
\bibitem[Go \latin{et~al.}(2017)Go, Hanke, Buhl, Freimuth, Bihlmayer, Lee,
  Mokrousov, and Bl{\ifmmode\ddot{u}\else\"{u}\fi}gel]{Go2017Apr}
Go,~D.; Hanke,~J.-P.; Buhl,~P.~M.; Freimuth,~F.; Bihlmayer,~G.; Lee,~H.-W.;
  Mokrousov,~Y.; Bl{\ifmmode\ddot{u}\else\"{u}\fi}gel,~S. {Toward surface
  orbitronics: giant orbital magnetism from the orbital Rashba effect at the
  surface of sp-metals}. \emph{Sci. Rep.} \textbf{2017}, \emph{7}, 1--10\relax
\mciteBstWouldAddEndPuncttrue
\mciteSetBstMidEndSepPunct{\mcitedefaultmidpunct}
{\mcitedefaultendpunct}{\mcitedefaultseppunct}\relax
\EndOfBibitem
\bibitem[Go \latin{et~al.}(2021)Go, Jo, Gao, Ando, Bl\"ugel, Lee, and
  Mokrousov]{PhysRevB.103.L121113}
Go,~D.; Jo,~D.; Gao,~T.; Ando,~K.; Bl\"ugel,~S.; Lee,~H.-W.; Mokrousov,~Y.
  Orbital Rashba effect in a surface-oxidized Cu film. \emph{Phys. Rev. B}
  \textbf{2021}, \emph{103}, L121113\relax
\mciteBstWouldAddEndPuncttrue
\mciteSetBstMidEndSepPunct{\mcitedefaultmidpunct}
{\mcitedefaultendpunct}{\mcitedefaultseppunct}\relax
\EndOfBibitem
\bibitem[Yoda \latin{et~al.}(2015)Yoda, Yokoyama, and Murakami]{Yoda2015Jul}
Yoda,~T.; Yokoyama,~T.; Murakami,~S. {Current-induced Orbital and Spin
  Magnetizations in Crystals with Helical Structure}. \emph{Sci. Rep.}
  \textbf{2015}, \emph{5}, 1--7\relax
\mciteBstWouldAddEndPuncttrue
\mciteSetBstMidEndSepPunct{\mcitedefaultmidpunct}
{\mcitedefaultendpunct}{\mcitedefaultseppunct}\relax
\EndOfBibitem
\bibitem[Yoda \latin{et~al.}(2018)Yoda, Yokoyama, and Murakami]{Yoda2018Feb}
Yoda,~T.; Yokoyama,~T.; Murakami,~S. {Orbital Edelstein Effect as a
  Condensed-Matter Analog of Solenoids}. \emph{Nano Lett.} \textbf{2018},
  \emph{18}, 916--920\relax
\mciteBstWouldAddEndPuncttrue
\mciteSetBstMidEndSepPunct{\mcitedefaultmidpunct}
{\mcitedefaultendpunct}{\mcitedefaultseppunct}\relax
\EndOfBibitem
\bibitem[Go and Lee(2020)Go, and Lee]{PhysRevResearch.2.013177}
Go,~D.; Lee,~H.-W. Orbital torque: Torque generation by orbital current
  injection. \emph{Phys. Rev. Res.} \textbf{2020}, \emph{2}, 013177\relax
\mciteBstWouldAddEndPuncttrue
\mciteSetBstMidEndSepPunct{\mcitedefaultmidpunct}
{\mcitedefaultendpunct}{\mcitedefaultseppunct}\relax
\EndOfBibitem
\bibitem[Go \latin{et~al.}(2020)Go, Freimuth, Hanke, Xue, Gomonay, Lee,
  Bl\"ugel, Haney, Lee, and Mokrousov]{PhysRevResearch.2.033401}
Go,~D.; Freimuth,~F.; Hanke,~J.-P.; Xue,~F.; Gomonay,~O.; Lee,~K.-J.;
  Bl\"ugel,~S.; Haney,~P.~M.; Lee,~H.-W.; Mokrousov,~Y. Theory of
  current-induced angular momentum transfer dynamics in spin-orbit coupled
  systems. \emph{Phys. Rev. Res.} \textbf{2020}, \emph{2}, 033401\relax
\mciteBstWouldAddEndPuncttrue
\mciteSetBstMidEndSepPunct{\mcitedefaultmidpunct}
{\mcitedefaultendpunct}{\mcitedefaultseppunct}\relax
\EndOfBibitem
\bibitem[Lee \latin{et~al.}(2021)Lee, Go, Park, Jeong, Ko, Yun, Jo, Lee, Go,
  Oh, Kim, Park, Min, Koo, Lee, Lee, and Lee]{Lee2021Nov}
Lee,~D. \latin{et~al.}  {Orbital torque in magnetic bilayers}. \emph{Nat.
  Commun.} \textbf{2021}, \emph{12}, 1--8\relax
\mciteBstWouldAddEndPuncttrue
\mciteSetBstMidEndSepPunct{\mcitedefaultmidpunct}
{\mcitedefaultendpunct}{\mcitedefaultseppunct}\relax
\EndOfBibitem
\bibitem[Ding \latin{et~al.}(2022)Ding, Liang, Go, Yun, Xue, Liu, Becker, Yang,
  Du, Wang, Yang, Jakob, Kl\"aui, Mokrousov, and Yang]{PhysRevLett.128.067201}
Ding,~S.; Liang,~Z.; Go,~D.; Yun,~C.; Xue,~M.; Liu,~Z.; Becker,~S.; Yang,~W.;
  Du,~H.; Wang,~C.; Yang,~Y.; Jakob,~G.; Kl\"aui,~M.; Mokrousov,~Y.; Yang,~J.
  Observation of the Orbital Rashba-Edelstein Magnetoresistance. \emph{Phys.
  Rev. Lett.} \textbf{2022}, \emph{128}, 067201\relax
\mciteBstWouldAddEndPuncttrue
\mciteSetBstMidEndSepPunct{\mcitedefaultmidpunct}
{\mcitedefaultendpunct}{\mcitedefaultseppunct}\relax
\EndOfBibitem
\bibitem[Hayashi \latin{et~al.}(2023)Hayashi, Jo, Go, Gao, Haku, Mokrousov,
  Lee, and Ando]{Hayashi2023Feb}
Hayashi,~H.; Jo,~D.; Go,~D.; Gao,~T.; Haku,~S.; Mokrousov,~Y.; Lee,~H.-W.;
  Ando,~K. {Observation of long-range orbital transport and giant orbital
  torque}. \emph{Commun. Phys.} \textbf{2023}, \emph{6}, 1--9\relax
\mciteBstWouldAddEndPuncttrue
\mciteSetBstMidEndSepPunct{\mcitedefaultmidpunct}
{\mcitedefaultendpunct}{\mcitedefaultseppunct}\relax
\EndOfBibitem
\bibitem[Choi \latin{et~al.}(2023)Choi, Jo, Ko, Go, Kim, Park, Kim, Min, Choi,
  and Lee]{Choi2023Jul}
Choi,~Y.-G.; Jo,~D.; Ko,~K.-H.; Go,~D.; Kim,~K.-H.; Park,~H.~G.; Kim,~C.;
  Min,~B.-C.; Choi,~G.-M.; Lee,~H.-W. {Observation of the orbital Hall effect
  in a light metal Ti}. \emph{Nature} \textbf{2023}, \emph{619}, 52--56\relax
\mciteBstWouldAddEndPuncttrue
\mciteSetBstMidEndSepPunct{\mcitedefaultmidpunct}
{\mcitedefaultendpunct}{\mcitedefaultseppunct}\relax
\EndOfBibitem
\bibitem[Ding \latin{et~al.}(2020)Ding, Ross, Go, Baldrati, Ren, Freimuth,
  Becker, Kammerbauer, Yang, Jakob, Mokrousov, and
  Kl\"aui]{PhysRevLett.125.177201}
Ding,~S.; Ross,~A.; Go,~D.; Baldrati,~L.; Ren,~Z.; Freimuth,~F.; Becker,~S.;
  Kammerbauer,~F.; Yang,~J.; Jakob,~G.; Mokrousov,~Y.; Kl\"aui,~M. Harnessing
  Orbital-to-Spin Conversion of Interfacial Orbital Currents for Efficient
  Spin-Orbit Torques. \emph{Phys. Rev. Lett.} \textbf{2020}, \emph{125},
  177201\relax
\mciteBstWouldAddEndPuncttrue
\mciteSetBstMidEndSepPunct{\mcitedefaultmidpunct}
{\mcitedefaultendpunct}{\mcitedefaultseppunct}\relax
\EndOfBibitem
\bibitem[Kim \latin{et~al.}(2021)Kim, Go, Tsai, Jo, Kondou, Lee, and
  Otani]{PhysRevB.103.L020407}
Kim,~J.; Go,~D.; Tsai,~H.; Jo,~D.; Kondou,~K.; Lee,~H.-W.; Otani,~Y. Nontrivial
  torque generation by orbital angular momentum injection in
  ferromagnetic-metal/$\mathrm{Cu}/{\mathrm{Al}}_{2}{\mathrm{O}}_{3}$
  trilayers. \emph{Phys. Rev. B} \textbf{2021}, \emph{103}, L020407\relax
\mciteBstWouldAddEndPuncttrue
\mciteSetBstMidEndSepPunct{\mcitedefaultmidpunct}
{\mcitedefaultendpunct}{\mcitedefaultseppunct}\relax
\EndOfBibitem
\bibitem[Krishnia \latin{et~al.}(2023)Krishnia, Sassi, Ajejas, Sebe, Reyren,
  Collin, Denneulin, Kovács, Dunin-Borkowski, Fert, George, Cros, and
  Jaffrès]{acs.nanolett.2c05091}
Krishnia,~S.; Sassi,~Y.; Ajejas,~F.; Sebe,~N.; Reyren,~N.; Collin,~S.;
  Denneulin,~T.; Kovács,~A.; Dunin-Borkowski,~R.~E.; Fert,~A.; George,~J.-M.;
  Cros,~V.; Jaffrès,~H. Large Interfacial Rashba Interaction Generating Strong
  Spin–Orbit Torques in Atomically Thin Metallic Heterostructures. \emph{Nano
  Letters} \textbf{2023}, \emph{23}, 6785--6791\relax
\mciteBstWouldAddEndPuncttrue
\mciteSetBstMidEndSepPunct{\mcitedefaultmidpunct}
{\mcitedefaultendpunct}{\mcitedefaultseppunct}\relax
\EndOfBibitem
\bibitem[Perdew \latin{et~al.}(1996)Perdew, Burke, and
  Ernzerhof]{PhysRevLett.77.3865}
Perdew,~J.~P.; Burke,~K.; Ernzerhof,~M. Generalized Gradient Approximation Made
  Simple. \emph{Phys. Rev. Lett.} \textbf{1996}, \emph{77}, 3865--3868\relax
\mciteBstWouldAddEndPuncttrue
\mciteSetBstMidEndSepPunct{\mcitedefaultmidpunct}
{\mcitedefaultendpunct}{\mcitedefaultseppunct}\relax
\EndOfBibitem
\bibitem[Kresse and Furthm\"uller(1996)Kresse, and
  Furthm\"uller]{PhysRevB.54.11169}
Kresse,~G.; Furthm\"uller,~J. Efficient iterative schemes for ab initio
  total-energy calculations using a plane-wave basis set. \emph{Phys. Rev. B}
  \textbf{1996}, \emph{54}, 11169--11186\relax
\mciteBstWouldAddEndPuncttrue
\mciteSetBstMidEndSepPunct{\mcitedefaultmidpunct}
{\mcitedefaultendpunct}{\mcitedefaultseppunct}\relax
\EndOfBibitem
\bibitem[Kresse and Joubert(1999)Kresse, and Joubert]{PhysRevB.59.1758}
Kresse,~G.; Joubert,~D. From ultrasoft pseudopotentials to the projector
  augmented-wave method. \emph{Phys. Rev. B} \textbf{1999}, \emph{59},
  1758--1775\relax
\mciteBstWouldAddEndPuncttrue
\mciteSetBstMidEndSepPunct{\mcitedefaultmidpunct}
{\mcitedefaultendpunct}{\mcitedefaultseppunct}\relax
\EndOfBibitem
\bibitem[Monkhorst and Pack(1976)Monkhorst, and Pack]{PhysRevB.13.5188}
Monkhorst,~H.~J.; Pack,~J.~D. Special points for Brillouin-zone integrations.
  \emph{Phys. Rev. B} \textbf{1976}, \emph{13}, 5188--5192\relax
\mciteBstWouldAddEndPuncttrue
\mciteSetBstMidEndSepPunct{\mcitedefaultmidpunct}
{\mcitedefaultendpunct}{\mcitedefaultseppunct}\relax
\EndOfBibitem
\bibitem[Liechtenstein \latin{et~al.}(1995)Liechtenstein, Anisimov, and
  Zaanen]{PhysRevB.52.R5467}
Liechtenstein,~A.~I.; Anisimov,~V.~I.; Zaanen,~J. Density-functional theory and
  strong interactions: Orbital ordering in Mott-Hubbard insulators. \emph{Phys.
  Rev. B} \textbf{1995}, \emph{52}, R5467--R5470\relax
\mciteBstWouldAddEndPuncttrue
\mciteSetBstMidEndSepPunct{\mcitedefaultmidpunct}
{\mcitedefaultendpunct}{\mcitedefaultseppunct}\relax
\EndOfBibitem
\bibitem[Dudarev \latin{et~al.}(1998)Dudarev, Botton, Savrasov, Humphreys, and
  Sutton]{PhysRevB.57.1505}
Dudarev,~S.~L.; Botton,~G.~A.; Savrasov,~S.~Y.; Humphreys,~C.~J.; Sutton,~A.~P.
  Electron-energy-loss spectra and the structural stability of nickel oxide: An
  LSDA+U study. \emph{Phys. Rev. B} \textbf{1998}, \emph{57}, 1505--1509\relax
\mciteBstWouldAddEndPuncttrue
\mciteSetBstMidEndSepPunct{\mcitedefaultmidpunct}
{\mcitedefaultendpunct}{\mcitedefaultseppunct}\relax
\EndOfBibitem
\bibitem[Grechnev \latin{et~al.}(2007)Grechnev, Di~Marco, Katsnelson,
  Lichtenstein, Wills, and Eriksson]{PhysRevB.76.035107}
Grechnev,~A.; Di~Marco,~I.; Katsnelson,~M.~I.; Lichtenstein,~A.~I.; Wills,~J.;
  Eriksson,~O. Theory of bulk and surface quasiparticle spectra for Fe, Co, and
  Ni. \emph{Phys. Rev. B} \textbf{2007}, \emph{76}, 035107\relax
\mciteBstWouldAddEndPuncttrue
\mciteSetBstMidEndSepPunct{\mcitedefaultmidpunct}
{\mcitedefaultendpunct}{\mcitedefaultseppunct}\relax
\EndOfBibitem
\bibitem[Cadi-Essadek \latin{et~al.}(2021)Cadi-Essadek, Roldan,
  Santos-Carballal, Ngoepe, Claeys, and de~Leeuw]{Cadi-Essadek2021}
Cadi-Essadek,~A.; Roldan,~A.; Santos-Carballal,~D.; Ngoepe,~P.~E.; Claeys,~M.;
  de~Leeuw,~N.~H. {DFT+U Study of the Electronic, Magnetic and Mechanical
  Properties of Co, CoO, and Co3O4}. \emph{S. Afr. J. Chem.} \textbf{2021},
  \emph{74}, 8--16\relax
\mciteBstWouldAddEndPuncttrue
\mciteSetBstMidEndSepPunct{\mcitedefaultmidpunct}
{\mcitedefaultendpunct}{\mcitedefaultseppunct}\relax
\EndOfBibitem
\bibitem[Bruno(1989)]{PhysRevB.39.865}
Bruno,~P. Tight-binding approach to the orbital magnetic moment and
  magnetocrystalline anisotropy of transition-metal monolayers. \emph{Phys.
  Rev. B} \textbf{1989}, \emph{39}, 865--868\relax
\mciteBstWouldAddEndPuncttrue
\mciteSetBstMidEndSepPunct{\mcitedefaultmidpunct}
{\mcitedefaultendpunct}{\mcitedefaultseppunct}\relax
\EndOfBibitem
\bibitem[Wang \latin{et~al.}(1993)Wang, Wu, and Freeman]{PhysRevB.47.14932}
Wang,~D.-s.; Wu,~R.; Freeman,~A.~J. First-principles theory of surface
  magnetocrystalline anisotropy and the diatomic-pair model. \emph{Phys. Rev.
  B} \textbf{1993}, \emph{47}, 14932--14947\relax
\mciteBstWouldAddEndPuncttrue
\mciteSetBstMidEndSepPunct{\mcitedefaultmidpunct}
{\mcitedefaultendpunct}{\mcitedefaultseppunct}\relax
\EndOfBibitem
\bibitem[van~der Laan(1998)]{GerritvanderLaan1998}
van~der Laan,~G. Microscopic origin of magnetocrystalline anisotropy in
  transition metal thin films. \emph{Journal of Physics: Condensed Matter}
  \textbf{1998}, \emph{10}, 3239\relax
\mciteBstWouldAddEndPuncttrue
\mciteSetBstMidEndSepPunct{\mcitedefaultmidpunct}
{\mcitedefaultendpunct}{\mcitedefaultseppunct}\relax
\EndOfBibitem
\bibitem[Dunn(1961)]{Dunn1961Jan}
Dunn,~T.~M. {Spin-orbit coupling in the first and second transition series}.
  \emph{Trans. Faraday Soc.} \textbf{1961}, \emph{57}, 1441--1444\relax
\mciteBstWouldAddEndPuncttrue
\mciteSetBstMidEndSepPunct{\mcitedefaultmidpunct}
{\mcitedefaultendpunct}{\mcitedefaultseppunct}\relax
\EndOfBibitem
\bibitem[Cole and Garrett(1970)Cole, and Garrett]{Cole1970Aug}
Cole,~G.~M.,~Jr.; Garrett,~B.~B. {Atomic and molecular spin-orbit coupling
  constants for 3d transition metal ions}. \emph{Inorg. Chem.} \textbf{1970},
  \emph{9}, 1898--1902\relax
\mciteBstWouldAddEndPuncttrue
\mciteSetBstMidEndSepPunct{\mcitedefaultmidpunct}
{\mcitedefaultendpunct}{\mcitedefaultseppunct}\relax
\EndOfBibitem
\bibitem[Xing \latin{et~al.}(2022)Xing, Miura, and Tadano]{PhysRevB.105.104427}
Xing,~G.; Miura,~Y.; Tadano,~T. Lattice dynamics and its effects on
  magnetocrystalline anisotropy energy of pristine and hole-doped
  ${\mathrm{YCo}}_{5}$ from first principles. \emph{Phys. Rev. B}
  \textbf{2022}, \emph{105}, 104427\relax
\mciteBstWouldAddEndPuncttrue
\mciteSetBstMidEndSepPunct{\mcitedefaultmidpunct}
{\mcitedefaultendpunct}{\mcitedefaultseppunct}\relax
\EndOfBibitem
\bibitem[Vajna \latin{et~al.}(2012)Vajna, Simon, Szilva, Palotas, Ujfalussy,
  and Szunyogh]{PhysRevB.85.075404}
Vajna,~S.; Simon,~E.; Szilva,~A.; Palotas,~K.; Ujfalussy,~B.; Szunyogh,~L.
  Higher-order contributions to the Rashba-Bychkov effect with application to
  the Bi/Ag(111) surface alloy. \emph{Phys. Rev. B} \textbf{2012}, \emph{85},
  075404\relax
\mciteBstWouldAddEndPuncttrue
\mciteSetBstMidEndSepPunct{\mcitedefaultmidpunct}
{\mcitedefaultendpunct}{\mcitedefaultseppunct}\relax
\EndOfBibitem
\bibitem[Salemi \latin{et~al.}(2019)Salemi, Berritta, Nandy, and
  Oppeneer]{Salemi2019Nov}
Salemi,~L.; Berritta,~M.; Nandy,~A.~K.; Oppeneer,~P.~M. {Orbitally dominated
  Rashba-Edelstein effect in noncentrosymmetric antiferromagnets}. \emph{Nat.
  Commun.} \textbf{2019}, \emph{10}, 1--10\relax
\mciteBstWouldAddEndPuncttrue
\mciteSetBstMidEndSepPunct{\mcitedefaultmidpunct}
{\mcitedefaultendpunct}{\mcitedefaultseppunct}\relax
\EndOfBibitem
\bibitem[Salemi \latin{et~al.}(2021)Salemi, Berritta, and
  Oppeneer]{PhysRevMaterials.5.074407}
Salemi,~L.; Berritta,~M.; Oppeneer,~P.~M. Quantitative comparison of
  electrically induced spin and orbital polarizations in heavy-metal/$3d$-metal
  bilayers. \emph{Phys. Rev. Mater.} \textbf{2021}, \emph{5}, 074407\relax
\mciteBstWouldAddEndPuncttrue
\mciteSetBstMidEndSepPunct{\mcitedefaultmidpunct}
{\mcitedefaultendpunct}{\mcitedefaultseppunct}\relax
\EndOfBibitem
\bibitem[Kleiner(1966)]{PhysRev.142.318}
Kleiner,~W.~H. Space-Time Symmetry of Transport Coefficients. \emph{Phys. Rev.}
  \textbf{1966}, \emph{142}, 318--326\relax
\mciteBstWouldAddEndPuncttrue
\mciteSetBstMidEndSepPunct{\mcitedefaultmidpunct}
{\mcitedefaultendpunct}{\mcitedefaultseppunct}\relax
\EndOfBibitem
\bibitem[Seemann \latin{et~al.}(2015)Seemann, K\"odderitzsch, Wimmer, and
  Ebert]{PhysRevB.92.155138}
Seemann,~M.; K\"odderitzsch,~D.; Wimmer,~S.; Ebert,~H. Symmetry-imposed shape
  of linear response tensors. \emph{Phys. Rev. B} \textbf{2015}, \emph{92},
  155138\relax
\mciteBstWouldAddEndPuncttrue
\mciteSetBstMidEndSepPunct{\mcitedefaultmidpunct}
{\mcitedefaultendpunct}{\mcitedefaultseppunct}\relax
\EndOfBibitem
\bibitem[Marzari and Vanderbilt(1997)Marzari, and
  Vanderbilt]{PhysRevB.56.12847}
Marzari,~N.; Vanderbilt,~D. Maximally localized generalized Wannier functions
  for composite energy bands. \emph{Phys. Rev. B} \textbf{1997}, \emph{56},
  12847--12865\relax
\mciteBstWouldAddEndPuncttrue
\mciteSetBstMidEndSepPunct{\mcitedefaultmidpunct}
{\mcitedefaultendpunct}{\mcitedefaultseppunct}\relax
\EndOfBibitem
\bibitem[Marzari \latin{et~al.}(2012)Marzari, Mostofi, Yates, Souza, and
  Vanderbilt]{RevModPhys.84.1419}
Marzari,~N.; Mostofi,~A.~A.; Yates,~J.~R.; Souza,~I.; Vanderbilt,~D. Maximally
  localized Wannier functions: Theory and applications. \emph{Rev. Mod. Phys.}
  \textbf{2012}, \emph{84}, 1419--1475\relax
\mciteBstWouldAddEndPuncttrue
\mciteSetBstMidEndSepPunct{\mcitedefaultmidpunct}
{\mcitedefaultendpunct}{\mcitedefaultseppunct}\relax
\EndOfBibitem
\bibitem[Wang \latin{et~al.}(2006)Wang, Yates, Souza, and
  Vanderbilt]{PhysRevB.74.195118}
Wang,~X.; Yates,~J.~R.; Souza,~I.; Vanderbilt,~D. Ab initio calculation of the
  anomalous Hall conductivity by Wannier interpolation. \emph{Phys. Rev. B}
  \textbf{2006}, \emph{74}, 195118\relax
\mciteBstWouldAddEndPuncttrue
\mciteSetBstMidEndSepPunct{\mcitedefaultmidpunct}
{\mcitedefaultendpunct}{\mcitedefaultseppunct}\relax
\EndOfBibitem
\bibitem[Ryoo \latin{et~al.}(2019)Ryoo, Park, and Souza]{PhysRevB.99.235113}
Ryoo,~J.~H.; Park,~C.-H.; Souza,~I. Computation of intrinsic spin Hall
  conductivities from first principles using maximally localized Wannier
  functions. \emph{Phys. Rev. B} \textbf{2019}, \emph{99}, 235113\relax
\mciteBstWouldAddEndPuncttrue
\mciteSetBstMidEndSepPunct{\mcitedefaultmidpunct}
{\mcitedefaultendpunct}{\mcitedefaultseppunct}\relax
\EndOfBibitem
\bibitem[Qiao \latin{et~al.}(2018)Qiao, Zhou, Yuan, and
  Zhao]{PhysRevB.98.214402}
Qiao,~J.; Zhou,~J.; Yuan,~Z.; Zhao,~W. Calculation of intrinsic spin Hall
  conductivity by Wannier interpolation. \emph{Phys. Rev. B} \textbf{2018},
  \emph{98}, 214402\relax
\mciteBstWouldAddEndPuncttrue
\mciteSetBstMidEndSepPunct{\mcitedefaultmidpunct}
{\mcitedefaultendpunct}{\mcitedefaultseppunct}\relax
\EndOfBibitem
\end{mcitethebibliography}

\pagebreak

\pagebreak

\pagebreak
\begin{figure}
\includegraphics[width=1.0\textwidth]{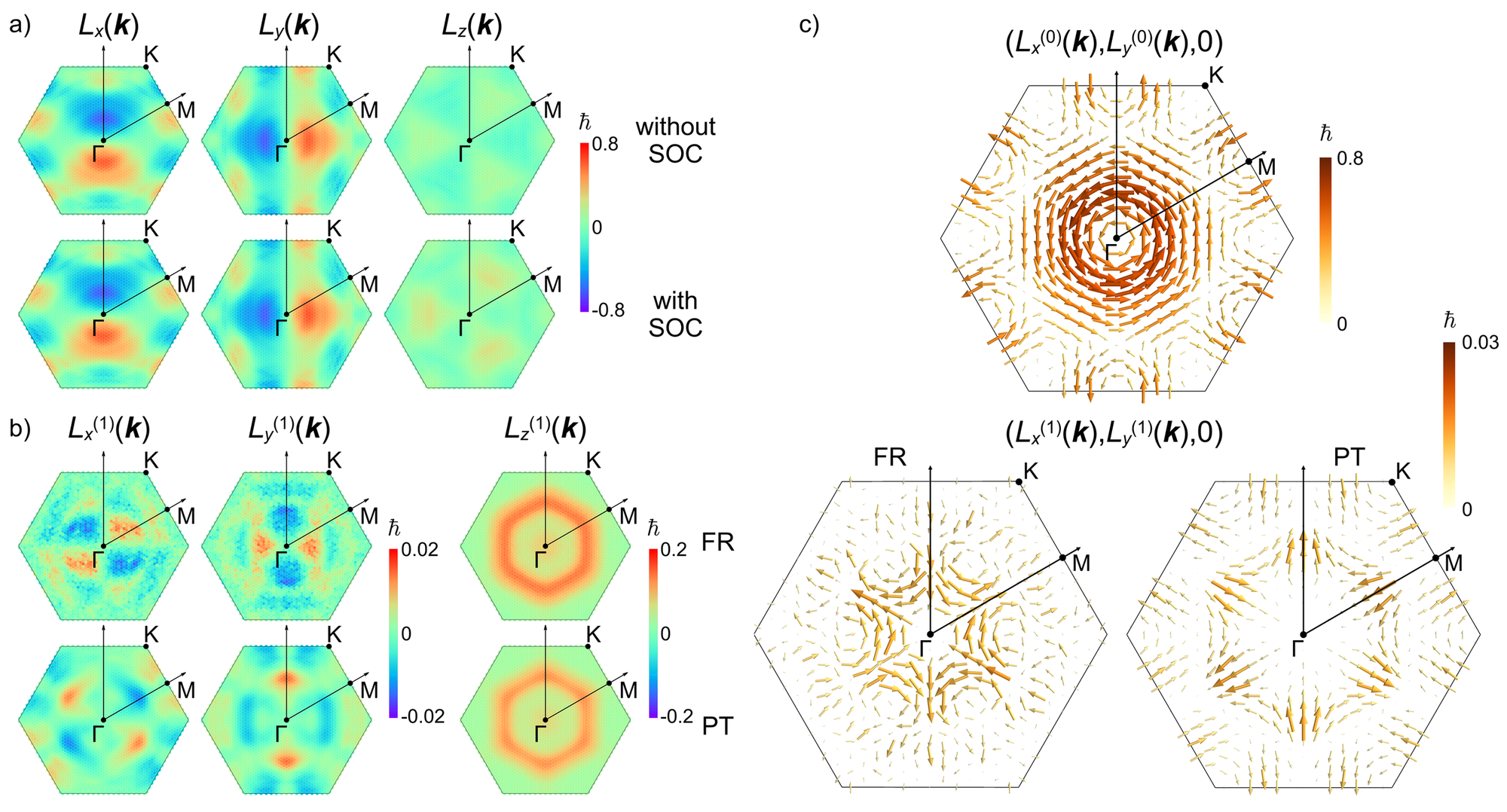}
\caption{a) Orbital profiles $\boldsymbol{L}(\boldsymbol{k})$ at the interfacial Co layer in 12Co(0001)/12Al(111) heterostructure as obtained from electronic structure calculations with and without spin-orbit coupling. b) Orbital profile $\boldsymbol{L}^{(1)}(\boldsymbol{k})$ induced by spin-orbit coupling as obtained from fully relativistic electronic structure calculations (FR) and perturbation theory (PT). c) In-plane orbital textures $\boldsymbol{L}^{(0)}(\boldsymbol{k})$ and $\boldsymbol{L}^{(1)}(\boldsymbol{k})$.}
\label{pic_4}
\end{figure}

\pagebreak

\end{document}